\documentclass[sigconf]{acmart}

\AtBeginDocument{%
}

\setcopyright{rightsretained}
\copyrightyear{2026}
\acmYear{2026}
\setcopyright{cc}
\setcctype{by}
\acmConference[UIST Adjunct '26]{The 39th Annual ACM Symposium on User Interface Software and Technology}{November 02--05, 2026}{Detroit, MI, USA}
\acmBooktitle{The 39th Annual ACM Symposium on User Interface Software and Technology (UIST Adjunct '26), November 02--05, 2026, Detroit, MI, USA}
\acmDOI{10.1145/3830397.3842492}
\acmISBN{979-8-4007-2855-6/2026/11}

\usepackage{xcolor}
\usepackage{svg}
\usepackage{caption}

\usepackage{listings}
\usepackage{xcolor}
\usepackage{seqsplit}
\begin{document}

\title{Demonstration of MANA: Multi-perspective Alignment of NPC Agents}

\author{Amir Tahmasbi}
\email{atahmasb@purdue.edu}
\affiliation{%
  \institution{Purdue University}
  \city{West Lafayette}
  \state{Indiana}
  \country{USA}
}

\author{Keshav Singh}
\email{sing1532@purdue.edu}
\affiliation{%
  \institution{Purdue University}
  \city{West Lafayette}
  \state{Indiana}
  \country{USA}
}

\author{Shaheer Raza}
\email{raza15@purdue.edu}
\affiliation{%
  \institution{Purdue University}
  \city{West Lafayette}
  \state{Indiana}
  \country{USA}
}

\author{Owen Mei}
\email{omei@purdue.edu}
\affiliation{%
  \institution{Purdue University}
  \city{West Lafayette}
  \state{Indiana}
  \country{USA}
}

\author{Jason Wu}
\email{jasonwu@purdue.edu}
\affiliation{%
  \institution{Purdue University}
  \city{West Lafayette}
  \state{Indiana}
  \country{USA}
}

\begin{abstract}
We present MANA (Multi-perspective Alignment of NPC Agents), an interactive approach for authoring NPC behavior through embodied demonstration. We show MANA in the context of Minecraft, where a developer aligns NPCs while playing the game itself. Activating MANA lets the developer dynamically switch into the point of view (POV) of any character and demonstrate, in-situ, how it would act given the surrounding context, through both movement and dialogue. MANA then distills each demonstration into a generalized policy that maps in-game context to actions. Finally, once the developer switches to another NPC or exits MANA mode, this generalized policy drives the NPC on its own. We demonstrate MANA by aligning two initially generic Minecraft NPCs: Fletcher and Farmer to act aggressively and cowardly, respectively. After roughly 2 minutes of in-game play, each character exhibits consistent and believable behaviors across new, unseen scenarios.
\end{abstract}

%

\ccsdesc[500]{Human-centered computing~Interactive systems and tools}

\keywords{Programming by Demonstration, NPC behavior authoring, embodied interaction, large language models, game development tools}
\begin{teaserfigure}
\captionsetup{skip=3pt}
  \centering
  \includegraphics[width=\textwidth]{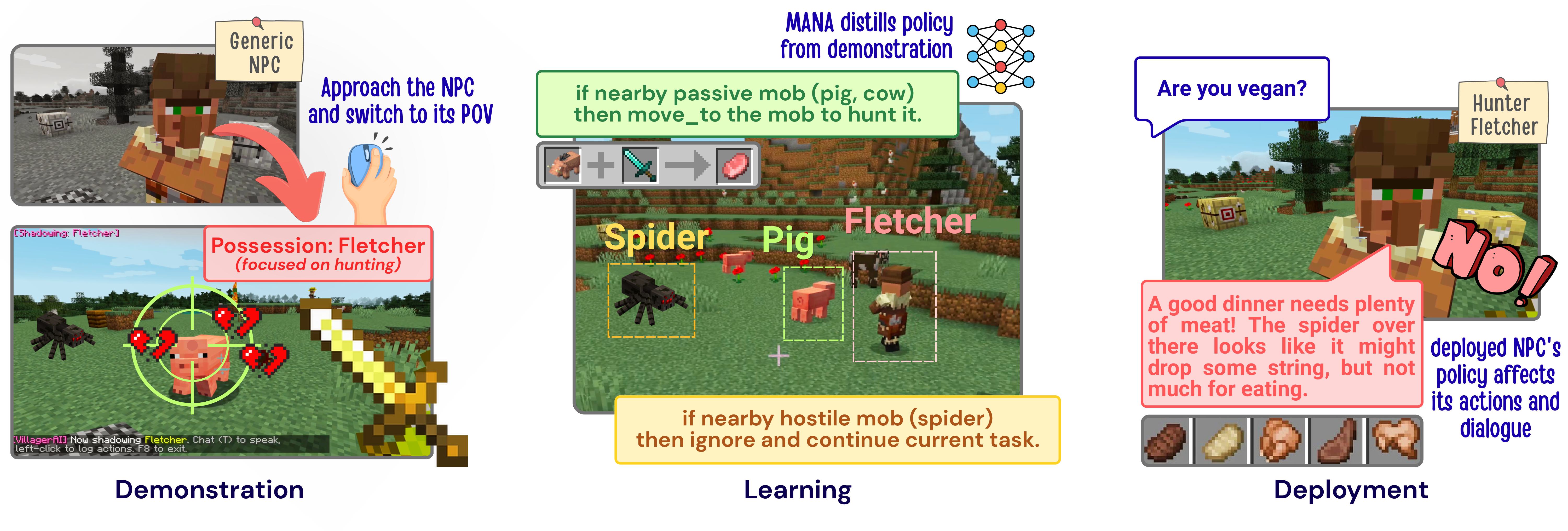}
  \caption{We implement MANA in Minecraft to align NPC behavior. The developer first takes over
the NPC's body to demonstrate contextual behavior
(\textit{Demonstration}). MANA distills the demonstration into a generalizable
policy (\textit{Learning}). The NPC then acts and speaks automatically, driven by its
learned profile and policy (\textit{Deployment}). Here, an in-game NPC, Fletcher, is aligned to hunt, fill its inventory with meat, provide persona-relevant dialogue.}
  \label{fig:teaser}
\end{teaserfigure}

\maketitle
\section{Introduction}

Non-Player Characters (NPCs) serve important roles in games where they are often necessary for game progression and enrich the gameplay experience.
Traditionally, developers defined NPC behavior through manually-authored logic (e.g., behavior trees and goal-oriented action planners)~\cite{orkin2006fear, yannakakis2018aigames}. However, it is difficult to anticipate a wide-range of in-game contexts and explicitly specify behavior for them all.
To this end, LLMs have contributed to (i) development-time authoring of NPC scripts~\cite{ubisoft2023ghostwriter} and (ii) supporting the creation of fully AI-driven characters~\cite{park2023generative,wang2023voyager}.
While fully AI-driven NPCs can dynamically generate a wider range of behaviors and create more immersive experiences, it is difficult for developers to define their behaviors.
Previous research found that end-users often struggle to author~\cite{zamfirescu2023johnny} and iterate through prompts~\cite{park2025visionary}.

To address this challenge, programming-by-demonstration (PBD) approaches are applied in domains such as interface design~\cite{myers1992demonstrational} and robotics~\cite{argall2009survey} as a way to specify behavior through examples rather than code.
PBD has also been explored for NPCs~\cite{drake2016demonstration,mehta2009authoring,sagredoolivenza2019trained}, but prior approaches often require separate training scenarios and offer limited generalization beyond the demonstrated situations.
We introduce MANA (Multi-perspective Alignment of NPC Agents), which lets developers inhabit multiple NPCs and demonstrate behavior as relevant contexts naturally arise during gameplay.
MANA distills each demonstration into a character-specific policy that preserves distinct personas across new situations, actions, and dialogue.

\section{MANA Implementation}

We implement MANA as a Minecraft mod, working in three phases: \emph{Demonstration}, \emph{Learning}, and \emph{Deployment}.


\textbf{Demonstration.} The developer walks up to any NPC and selects the \textit{possession} option to assume control of their POV. In this mode, the developer can choose to move, fight, build, and talk as the character, while the system logs everything in the background. Formally, a demonstration is a pair $\mathcal{D}=(\mathcal{U},\mathcal{E})$: $\mathcal{U}$ is the speech stream, dialogue exchanged in general chat or by approaching other NPCs and opening a chat-box with them; $\mathcal{E}$ is the embodied record, logged every $T$ ticks, pairing each action performed with the neighborhood around the agent at that moment,
\[
  \mathcal{E}=\big\langle (a,\mathcal{N}) \big\rangle,\qquad
  a\in\mathcal{A},\quad \mathcal{N}=\{(c,d,\rho)\},
\]
where $a$ is a primitive action from the vocabulary $\mathcal{A}$ (e.g.\ \texttt{place\_block}, \texttt{hit\_entity}, \texttt{move\_to}), and $\mathcal{N}$ lists nearby mobs, items, and NPCs, each with a direction $d$ (e.g.\ \texttt{"N"}) and a distance $\rho$ in blocks.

\textbf{Learning.} Once the developer releases the NPC, an LLM reviews the recorded speech and actions and distills them into a behavior profile $\mathcal{B}=\Sigma(\mathcal{D})=(\mathcal{V},\mathcal{P})$: $\mathcal{V}$ is a natural-language persona capturing the character's personality and tone, and $\mathcal{P}$ is an interpretable action policy, a set of goals governed by spatial, sequencing, and persistence rules that determine when the character attacks, ignores an entity, or builds.

\textbf{Deployment.} From here, the NPC acts on its own. At each moment it perceives an observation $\mathcal{O}$, its current neighborhood $\mathcal{N}$ together with world context, and responds through the profile-conditioned policy $\mathcal{A}=\pi(\mathcal{B},\mathcal{O})$, generating its own speech and invoking actions from $\mathcal{A}$, along with their arguments, to drive itself autonomously.
\section{Scenario 1: Battle}
The environment is populated with hostile mobs at night. Here we possess two NPCs and align them with opposing temperaments. We then observe the resulting behavior from their distinct profiles. \\
\textit{\textbf{Aggressive Fletcher.}} The developer possesses the Fletcher and fights zombies and creepers on sight while keeping distance from the player. Each engagement is logged every 20 ticks as an action key, e.g. \texttt{\seqsplit{"hit\_entity"}}, together with its  \texttt{\seqsplit{"target":"creeper"}}, and the surrounding neighborhood, giving each nearby entity's direction and distance relative to the agent, e.g. \texttt{\seqsplit{"dir":"NW","dist":12}} for the creeper and \texttt{\seqsplit{"dir":"E","dist":3}} for the player standing back. The language is also captured through the chat-box: the player opens a conversation and the developer improvises the Fletcher's dialogue in real time, For example, their response to the player's remark, \textit{``hunting creepers is so risky!''}, is \textit{``life is short dude! let's have fun, don't worry about that,''} which gives the model clues about their attitude toward life and their tone. \\
After demonstration, the model learns the persona and action policy. From the temporal logs, learning phase distills an ordered goal policy: a nearby hostile mob triggers a \texttt{\seqsplit{move\_to}} and \texttt{\seqsplit{attack}} response, while the player being nearby triggers \texttt{\seqsplit{move\_to}} within talking range; and from the dialogues, the resulting persona's \texttt{"tone"} is casual, confident, thrill-seeking, with slang word-pieces like ``dude!'' recurring in their responses. At deployment, the Fletcher profile drives the NPC autonomously: it hunts hostiles and narrates its own actions in voice, e.g. \textit{"There's a creeper nearby, bro. Time to get ready for some action and take it out before it blows up the place,"} or, on noticing the player, \textit{"Gotta stick close, bro, especially since we're in battle mode!"}

\textit{\textbf{Cowardly Farmer.}} Under the same threat conditions, the developer possesses the Farmer and performs the opposite response: as zombies approach, the Farmer flees, builds an enclosed shelter, and stays inside. The construction is logged with action key \texttt{\seqsplit{"place\_block"}}, with material \texttt{\seqsplit{"warped\_planks"}}, the time of day \texttt{"NIGHT"}, the agent's unchanged health \texttt{"hp":"20/20"}, and a hostile creeper still nearby. Learning phase produces an avoidance-first policy: a hostile entity nearby while building does not interrupt construction, and even a friendly villager approaching during a crisis is ignored in favor of continuing the primary survival task. Critically, the shelter itself becomes part of the profile. Each demonstrated block is stored by its relative offset and material under a named template, with the relative position of blocks to each other, and the shelter's world $xyz$ coordinates are saved alongside it. This lets the deployed Farmer rebuild the same shelter and return to the same location on subsequent nights rather than improvising a new one. At deployment, safety overrides any competing instruction: if the player orders it to \textit{``leave the shelter and play football with me,''} at night, the Farmer refuses and remains inside until the threat or during \texttt{"NIGHT"} state has passed.

\begin{figure}[t]
  \centering
      \includegraphics[width=\linewidth]{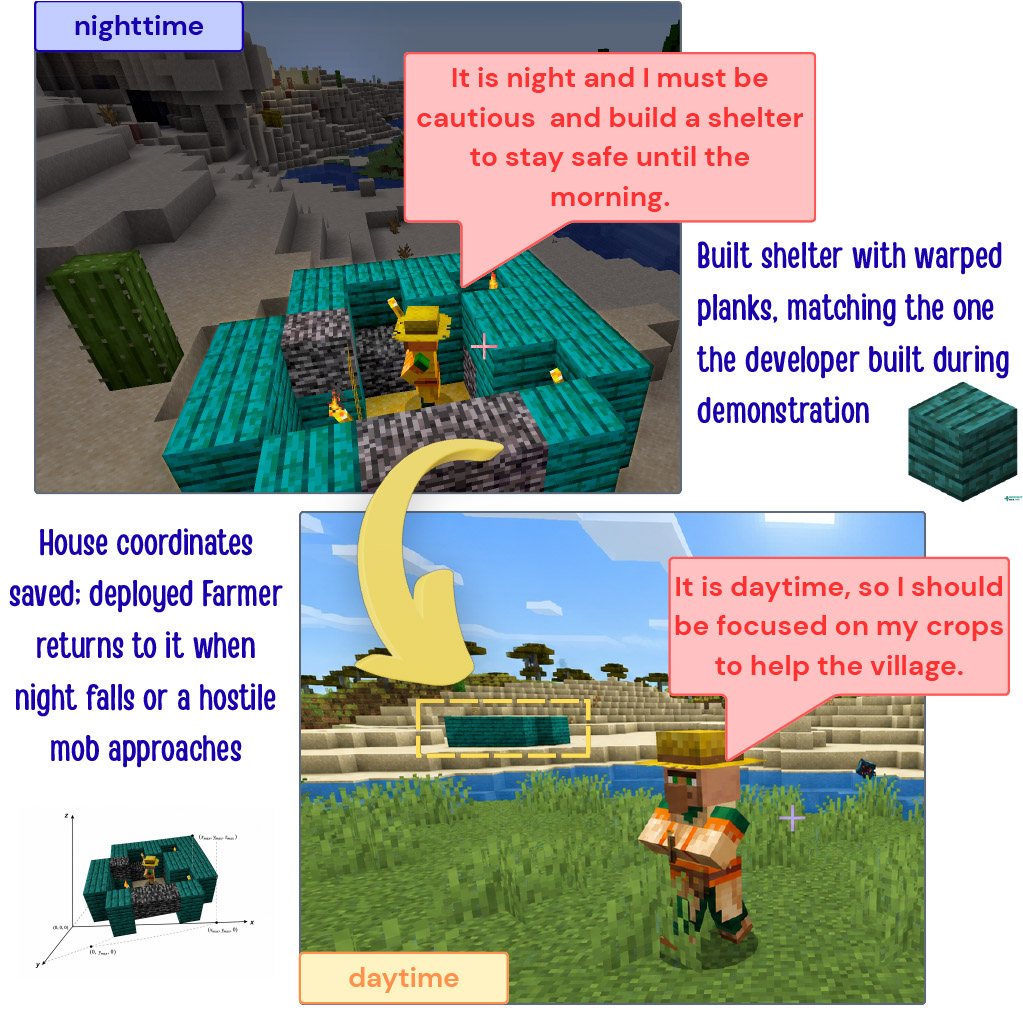}
  \caption{Deployed Cowardly Farmer behavior across day and night. At night, it returns to a shelter location learned during demonstration. During the day, it farms when no threat is present.}
  \label{fig:farmer-day-night}
\end{figure}
\section{Scenario 2: Socialization}
In this scenario, NPCs live through a peaceful, clear day doing their routines, occasionally talking to each other when the player prompts them with a task. One of our goals here is to show that, while language and action demonstration together make for a richer character than either alone, even a single modality can expand into the other.\\
The developer shapes the Fletcher NPC by having it hunt passive mobs like pigs in the environment, with each kill logged as an inventory change, e.g.\ \texttt{\seqsplit{"added":["porkchop:1"]}}, without ever using the chat-box channel to shape its persona. After demonstration, the Learning phase generates a policy in which a nearby passive mob, triggers a \texttt{\seqsplit{"move\_to"}} and \texttt{\seqsplit{"hit\_entity"}} response, while a nearby hostile mob, e.g.\ \texttt{"creeper, spider"}, is met with \texttt{"ignore and continue current task"}, showing the model distinguishing between mob types purely from action demonstration. Notably, even though language was never demonstrated, the Fletcher's actions still surface through language at deployment: asked what an ideal dinner for the party would be, it responds \textit{``I'd say a good dinner needs plenty of meat,''} and when asked whether it is vegan, it flatly says no, since it hunts.

These traits also surface in NPC-to-NPC interaction. We prompted our two shaped characters, the \textit{cowardly Farmer} and the \textit{hunter Fletcher}, to choose a gift for the player. The Farmer proposed bread and a torch, while Fletcher proposed a sturdy pickaxe or a sharp sword; the two ultimately converged on the torch, with the Farmer insisting on safety as its reasoning and Fletcher accepting the torch as practical for exploration and keeping shadows at bay.
Additionally, when asked to organize the party, the Farmer offered, \textit{``I can help with farming for party food, brother, no need for risky hunting,''} to which Fletcher replied, \textit{``No need for risky hunting? That's your way, Farmer, but a feast needs a good game!''} The Farmer concluded, \textit{``It's good we both contribute in our own ways, brother! I understand your need to hunt, but please be careful.''} The exchange reflects the two personas established in demonstrations: a cautious Farmer and a risk-taking Fletcher, projected here into an entirely new social context without any additional demonstration.


\bibliographystyle{ACM-Reference-Format}
\bibliography{sample-base}

\end{document}